\documentclass[sigconf,screen]{acmart}
\AtBeginDocument{%
  }

\copyrightyear{2026}
\acmYear{2026}
\setcopyright{cc}
\setcctype{by}
\acmConference[SIGGRAPH Real-Time Live! '26]{Special Interest Group on Computer Graphics and Interactive Techniques Conference Real-Time Live!}{July 19--23, 2026}{Los Angeles, CA, USA}
\acmBooktitle{Special Interest Group on Computer Graphics and Interactive Techniques Conference Real-Time Live! (SIGGRAPH Real-Time Live! '26), July 19--23, 2026, Los Angeles, CA, USA}
\acmDOI{10.1145/3799821.3820912}
\acmISBN{979-8-4007-2544-9/2026/07}

\allowdisplaybreaks 

\usepackage[labelfont=bf,format=plain]{caption}
\usepackage{subcaption}
\begin{document}

\title[HiPR: Hierarchical Progressive Rendering for Immediate Feedback]{HiPR: Hierarchical Progressive Rendering\\for Immediate Feedback}


\author{Rafael Padilla}
\affiliation{%
  \institution{University of Utah}
  \city{Salt Lake City, UT}
  \country{USA}}
\email{rpadiper@gmail.com}

\author{Andrew W. Tate}
\affiliation{%
  \institution{University of Utah}
  \city{Salt Lake City, UT}
  \country{USA}}
\email{andrew.tate@utah.edu}

\author{Dhruv Ram}
\affiliation{%
  \institution{University of Utah}
  \city{Salt Lake City, UT}
  \country{USA}}
\email{dhruvram.dr@gmail.com}

\author{Cem Yuksel}
\affiliation{%
  \institution{University of Utah}
  \city{Salt Lake City, UT}
  \country{USA}}
\email{cem@cemyuksel.com}

\renewcommand{\shortauthors}{Rafael Padilla et al.}  

\begin{abstract}
\emph{Hierarchical Progressive Rendering} (HiPR) is a dynamic render-scheduling algorithm that makes interactive path tracing finally feel real-time. While most renderers recompute the entire frame after any change to the scene, our method 
updates some of the pixels based on a priority order while keeping the others unchanged.
Rather than relying on error-driven or temporal reuse heuristics, it amortizes rendering costs by organizing pixels into a hierarchy of light-path dependencies from changed elements outward, prioritizing by perceptual impact and delivering instant visual feedback, while eventually converging to an unbiased result.
\end{abstract}


\begin{teaserfigure}
 \includegraphics[width=\linewidth,trim=0 60 0 60,clip]{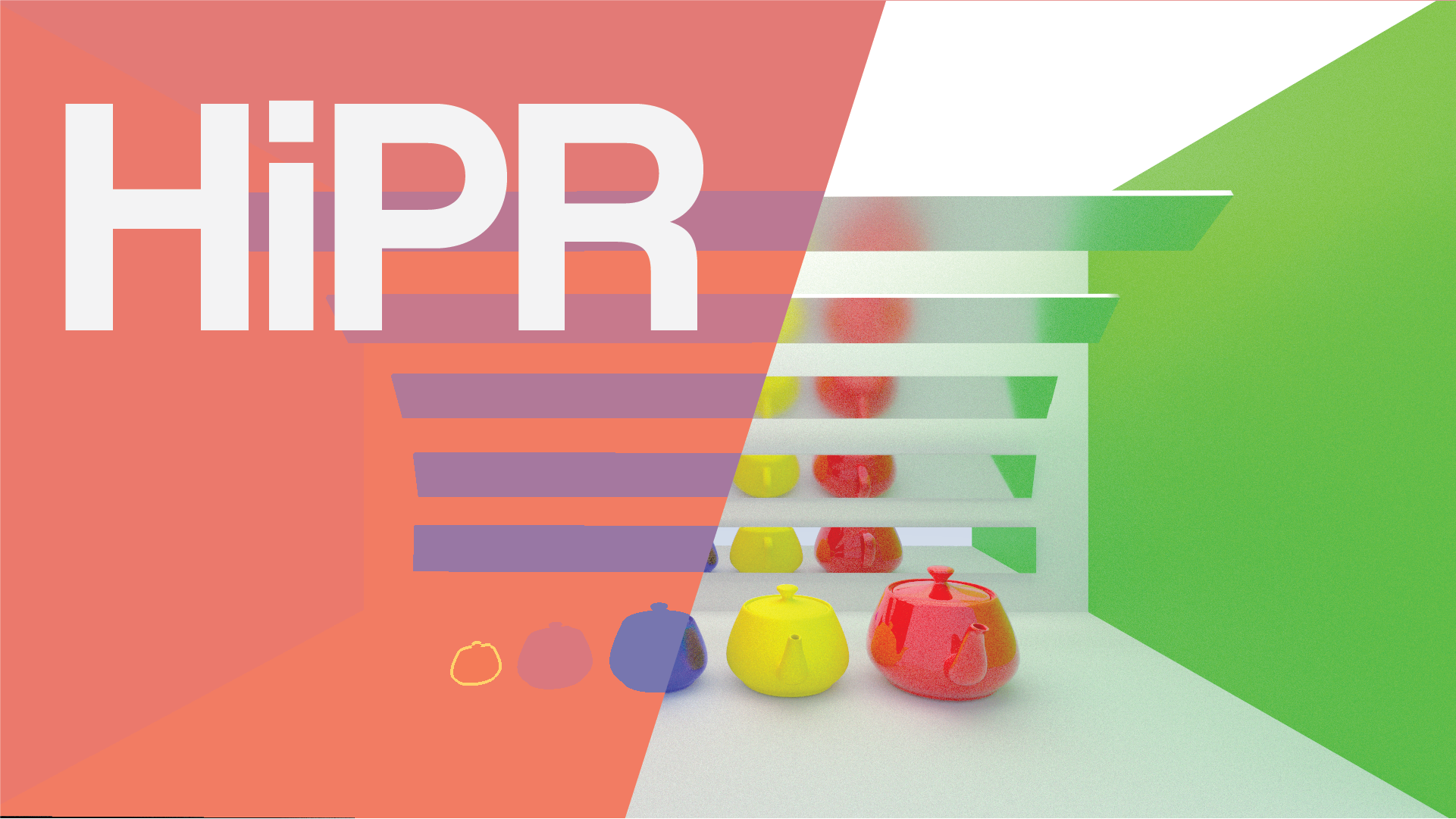}
  \caption{An example scene demonstrating our Hierarchical Progressive Rendering: (left)~object id visualization; (right)~final rendered image.}
  \label{fig:teaser}
\end{teaserfigure}


\maketitle

\section{Introduction}
Interactive path tracing remains 
prohibitively expensive
for scene editing applications,
from look development and virtual production to interactive design tools.
Existing methods such as adaptive sampling and spatiotemporal reuse amortize sample cost across frames. However, 
these methods are not designed for scene modifications, so
they do not identify which pixels reveal a scene change and schedule rendering accordingly. We present \emph{Hierarchical Progressive Rendering} (HiPR), a scheduling algorithm for rendering that takes a fundamentally different approach by leveraging the structure of light-path dependencies and perceptual salience to guide pixel priorities while preserving unbiased convergence.

\section{Hierarchical Progressive Rendering}

The goal of our method is to determine which pixels must be updated after a scene modification. We begin by splitting the framebuffer into a grid of tiles $\mathcal{T}$ of user-specified dimensions $n \times n$. For each scene modification, we find the tiles that are impacted (via either direct visibility or indirect light transport) and re-render them using a priority policy.

Our method works in three stages. In the first stage, we perform a visibility pass to determine which tiles are directly impacted by the modified scene data. In the second stage, we render those directly impacted tiles with path tracing. During this stage, we also determine which other tiles are impacted by secondary light bounces. At the end of the second stage, we order all tiles into a priority queue. In the final stage, we render the remaining tiles in the order of their priority.

\subsection{Stage 1: Initial Visibility Pass}

We define the \emph{changed set} $\mathcal{E}$ as the set of elements with modified transformation, geometry, material, or emission data as compared to the previous scene description. The first stage finds the tiles where elements in $\mathcal{E}$ are directly visible.

We use a visibility pass that traces one primary ray per pixel center and stores a G-buffer of object identity and camera-space depth $Z$. The active set of tiles $\mathcal{A}$ is seeded with every tile in which an element in $\mathcal{E}$ is visible.

\subsection{Stage 2: Render Directly Visible Tiles}

In the second stage, we render the tiles in set $\mathcal{A}$ using path tracing to give immediate visual feedback to the user. In order to discover the other tiles affected by secondary bounces, we find the corresponding tile for each secondary hit point $\mathbf{x}_k$ (i.e., a vertex along the light path), where $k$ is the bounce count. The world space $\mathbf{x}_k$ is transformed into camera space $[u\;v]^T$ with a projection function $P$, such that $[u\;v]^T=P(\mathbf{x}_k)$. This projection $P$ is defined by the inverse camera transform $M_{\mathrm{cam}}^{-1}$ using a pinhole model, such that
\begin{align}
  \begin{bmatrix} u \\ v \end{bmatrix} &=
  \begin{bmatrix} O_x + f \dfrac{x}{z} \\[1.0em]
                  O_y + f \dfrac{y}{z} \end{bmatrix}
&&\text{with}&
\begin{bmatrix} x \\ y \\ z \end{bmatrix}
  &= M_{\mathrm{cam}}^{-1}\,\mathbf{x}_k,
\end{align}
where $[O_x\;O_y]$ is the camera's principal point, $f$ is the focal length (square pixels are assumed). We write $d(\mathbf{x}_k) = z$ for its depth (i.e., camera-space z coordinate).

A projected vertex contributes to its tile only if it falls within the framebuffer, its depth is within a range $\epsilon$ of the stored depth value in the G-buffer, and it is otherwise ignored. The depth test is required to discern whether the point is visible.

When a vertex passes the depth test, it marks the tile as discovered and atomically increments the tile's weight. Since each vertex may have a different impact depending on light interaction type, this increment is computed using
\begin{equation}
  \Delta W(\mathbf{x}_k) =
  \beta_k \, w_{\mathrm{lobe}}(\ell_k)\, w_{\mathrm{path}}(c_k)\,
  \frac{1}{1+k},
  \label{eq:weight}
\end{equation}
where $\beta_k$ is the throughput accumulated to $\mathbf{x}_k$, and $\tfrac{1}{1+k}$ adds an explicit depth penalty atop throughput's natural falloff. The two remaining terms are user-defined priorities, applied independently of the sampler's importance distribution, and each forms a weighted sum over its categories: \begin{itemize}
\item $w_{\mathrm{lobe}}(\ell_k)$ ranks the lobe sampled at the vertex---diffuse, glossy, specular, transmission, with $\sum_\ell w_{\mathrm{lobe}}(\ell) = 1$; 
\item $w_{\mathrm{path}}(c_k)$ ranks the class of the path \emph{prefix}, so it is known mid-path, with $\sum_c w_{\mathrm{path}}(c) = 1$: direct ($LDE$, $LSE$), indirect ($LDD^{+}E$), or caustic ($LS^{+}DE$). Cast-shadow changes, which no scatter vertex reaches, are captured during next-event estimation: when a shadow ray toward a light is blocked by $\mathcal{E}$, the shadowed tile receives the occlusion priority.
\end{itemize}

A tile's weight accumulates the contributions of all vertices that project into it and pass the depth test using atomic summation
\begin{equation}
  W(T) = \sum_{\mathbf{x}_k \in T} \Delta W(\mathbf{x}_k).
  \label{eq:tile}
\end{equation}
We propose building a light transport hierarchy using tile weights $W$.

\subsection{Stage 3: Sort and Render Tiles}

After the tiles in $\mathcal{A}$ conclude rendering, we perform a parallel sort of the discovered tiles by weight. Then, we render in descending order.

HiPR only reorders tiles; it does not alter the per-sample estimator for a tile's pixels. Each pixel remains the mean of i.i.d. unbiased samples, so scheduling does not bias the final path tracing result. Tiles without discovered transport reuse the prior frame's radiance until they are rendered. Eventually, after the discovered tiles are rendered, we render the undiscovered tiles to ensure that the final result is unbiased.

\subsection{Parameters}

HiPR exposes a small set of parameters, each trading responsiveness against throughput:
\begin{itemize}
  \item $n$, the tile size: small tiles localize feedback and keep the depth test
  exact, but fragment GPU dispatch and enlarge the selection; large tiles are coherent
  and cheap, but coarsen discovery toward the identity-only test.
  \item $\epsilon$, the depth tolerance: tight values risk false negatives on grazing
  or curved surfaces; loose values admit false positives through thin geometry.
  \item $w_{\mathrm{lobe}}$ and $w_{\mathrm{path}}$, the perceptual priority weights:
  use-case specific. An artist lighting a glass interior raises the caustic and
  transmission weights to surface refracted focusing first; a character
  look-development pass favors diffuse and direct response; a mirror-heavy set favors
  the specular lobe.
\end{itemize}

As a sensible default, we use the order suggested by \citet{ulschmid2025automated} (specular and direct ahead of diffuse, with indirect and caustic
trailing), as it approximates the order in which the eye notices a change.

\section{Reference Implementation}
HiPR was implemented in a compact path-tracer as a reference, which was demonstrated during SIGGRAPH 2026 Real Time Live! Our renderer leverages GPU hardware acceleration, specifically ray tracing and BVH traversal cores exposed through the Vulkan API. Rather than adopting
the full Vulkan ray tracing pipeline, our renderer uses inline ray tracing exclusively, building
hardware acceleration structures and manually binding materials to object IDs for greater
flexibility and control. Materials are modeled using Adobe's
implementation of the OpenPBR BSDF, providing a physically grounded,
industry-standard shading model.


The standard Vulkan ray tracing pipeline couples acceleration structures with a shader
binding table that maps BLASes to their corresponding shaders. This design assumes multiple
shader stages and a flexible dispatch model; overhead that is unnecessary for a single megakernel
path tracer. Inline ray queries let us issue a ray and receive hit information, including an object
ID, directly in the compute shader, cutting out the pipeline machinery entirely. Our materials are
keyed to these object IDs, and their parameters are packed to satisfy GPU memory alignment
requirements, so the object ID alone is sufficient to resolve the full material state at any hit point.

\section{Applications}

While HiPR can be applied to any progressive path-tracing system, its greatest benefits arise in interactive workflows where users repeatedly modify scene content and require immediate visual feedback. Rather than treating every pixel as equally important after a scene edit, HiPR prioritizes the regions most likely to reveal the consequences of that change, enabling a more responsive editing and exploration experience. We highlight several representative applications below.

\subsection{Look Development}
Artists frequently explore multiple material, lighting, and composition alternatives before settling on a final look. Because these workflows are inherently iterative, the time between making a change and seeing its visual impact directly affects productivity. HiPR reduces this delay by immediately revealing the portions of the image most affected by an edit, enabling faster creative exploration and decision making.

\subsection{Virtual Production}
Virtual production workflows often involve multiple stakeholders evaluating a scene simultaneously. Directors, lighting artists, and technical artists may request adjustments to props, materials, or lighting while reviewing a shot in real time. By providing meaningful visual feedback before a full frame converges, HiPR helps maintain creative momentum during collaborative review sessions and live stage production.

\subsection{Attention-Guided Applications}
In interactive applications, users rarely distribute their attention uniformly across the screen. HiPR can exploit this observation by selecting objects of interest from game-play events, user interaction, or eye-tracking data. For instance, when a player opens a brightly lit doorway into dark room, HiPR could immediately prioritize the doorway, nearby surfaces, and the indirect illumination introduced into the environment. Likewise, a destructible wall, exploding vehicle, or other visually significant event could serve as the initial HiPR object, allowing the most perceptually important consequences of the event to appear first. In virtual and augmented reality applications, eye-tracking systems could similarly identify the user's object of focus, enabling rendering effort to propagate outward from the region receiving the most attention.

\section{Conclusion and Future Work}
We have presented HiPR, a dynamic render-scheduling algorithm which reduces the latency and provides visual feedback through a path-traced render following the change of a scene description.

The current formulation re-projects the entire frame after camera motion, yet parallax affects objects unequally: as the camera orbits, a distant background may shift by less than a pixel while a foreground character moves substantially. Per-object heuristics that decide whether camera motion warrants a fresh HiPR pass would let the renderer skip objects below a perceptual change threshold. It would also be interesting to explore delta-aware weighting: scaling priorities by the magnitude of the edit rather than the changed object's contribution alone. This would enable scheduling to track visible difference directly. Additionally, we see a possible use case in leveraging simulation data and motion vectors to further guide sample density in complex simulated 3D environments. Another future work direction would be establishing a perceptual light transport salience formula to avoid the need for manual parametrization for our perceptual priority weights.

In conclusion, we believe that HiPR reframes progressive rendering as a scheduling problem. By leveraging light-transport dependencies to determine what should be rendered next, HiPR delivers immediate feedback after scene edits and opens new opportunities for interactive path tracing in content creation, virtual production, and real-time experiences.

\bibliographystyle{ACM-Reference-Format}
\bibliography{sample-base}

\end{document}